\documentclass[aps,prl,twocolumn,showpacs,floatfix]{revtex4}
\pdfoutput=1

\usepackage{amssymb,amsmath}
\usepackage{graphicx}

\def\Dc{\Delta_{\mathrm{C}}}
\def\ud{\mathrm{d}}

\def\oA{\omega_{\mathrm{A}}}
\def\oR{\omega_{\mathrm{R}}}
\def\DA{\Delta_{\mathrm{A}}}
\def\DC{\Delta_{\mathrm{C}}}
\def\dt{\frac{\partial}{\partial t}}
\def\R{\vec{\hat{R}}}

\renewcommand{\Re}{\mathrm{Re}}
\renewcommand{\vec}[1]{\underline{#1}}
\newcommand{\mat}[1]{\mathbf{#1}}
\newcommand{\rev}[1]{\vec{r}^{(#1)}}
\newcommand{\lev}[1]{\vec{l}^{(#1)}}

\begin{document}

\title{Excess noise depletion of a Bose-Einstein condensate in an optical cavity}
\author{G.\ Szirmai}
\author{D.\ Nagy}
\author{P.\ Domokos}
\affiliation{Research Institute for Solid State Physics and Optics, Hungarian Academy of Sciences, H-1525 Budapest P.O. Box 49, Hungary}

\begin{abstract}
Quantum fluctuations of a cavity field coupled into the motion of ultracold bosons  can be strongly amplified by a mechanism analogous to the Petermann excess noise factor in lasers with unstable cavities. For a Bose-Einstein condensate in a stable optical resonator, the excess noise effect amounts to a significant depletion on long timescales.
\end{abstract}


\maketitle

A Bose-Einstein condensate (BEC) in a high-finesse optical resonator realizes a peculiar hybrid system of strongly coupled matter and light waves. The condensate with its large optical density creates  a substantial refractive medium for the light field, meanwhile the cavity field exerts significant mechanical forces on the motion of the atom cloud. 
The study of quantum properties arising from this nonlinear dynamics has just begun \cite{vukics07b,larson07b,maschler08,cola04a,zhang08}, accompanied by an experimental progress in loading ultracold atoms in the tiny volume of a high-finesse microcavity \cite{brennecke07,slama07,colombe07}.

Many of the properties of a degenerate quantum gas can be satisfactorily accounted for by assuming a `condensate wavefunction' which obeys a non-linear Schr\"odinger equation. Although the atom-atom collisions constantly kick out atoms from the one-particle ground state, even at zero temperature, the fraction of noncondensed atoms due to s-wave scattering is usually negligible. Large depletion of a BEC can  indicate some inherent many-body interaction effect. In case of collisions, such observation requires large local densities. An example is the strong depletion demonstrated as a precursor of the phase transition from superfluid to Mott state in an optical lattice \cite{stoferle04,xu06}. In this Letter we will show that a condensate dispersively coupled to the radiation field in a cavity can also be subjected to strong depletion.  In this case the many-body effect originates from the long-range atom-atom coupling mediated by the cavity mode, therefore a significant amount of depletion may occur at low density.


We consider an open system, a pumped and lossy cavity field interacting with a BEC inside. This  system has a dynamical equilibrium which self-consistently defines the condensate wavefunction.
Depletion of the BEC is defined as the total occupancy of motional modes other than the macroscopically populated  BEC wavefunction. For the atoms in a cavity, the quantum fluctuations of the radiation field is a possible source of depletion. However, such a fluctuating dipole potential alone cannot imply the unexpectedly large depletion in the dispersive interaction limit. As it was calculated for a probe field propagating freely through the condensate \cite{leonhardt99}, the depletion scales as the absorption which will be suppressed by choosing very large detuning. Therefore the cavity is essential in creating a specific coupling of the many-atom system to the radiation. Because of  the fast round-trips of photons, the cavity field experiences the \emph{collective} behavior of atoms.  This kind of  coupling gives rise to the noise amplification mechanism analogous to the so called `excess noise' in laser physics \cite{petermann89, hamel90,cheng96,eijkelenborg96}.  

The single-mode high-Q optical cavity is driven by injecting coherent laser light through one of its mirrors.  The laser frequency $\omega$ is detuned far from the frequency of the atomic transition $\oA$, i.e., $|\DA|\gg\gamma$, with the detuning $\DA\equiv\omega-\oA$, and $2\gamma$ being the rate of spontaneous emission. On the other hand, the laser is close to resonance with the cavity. A one-dimensional geometry is considered for the atomic motion. The single mode function of the cavity field is given by $\cos(k x)$, $k=\omega/c$. The dynamics of its annihilation operator is governed by the following Heisenberg-Langevin-type equation of motion in the rotating wave approximation:
\begin{subequations}
\label{eqmo}
\begin{multline}
\label{eqmorad}
i\dt\hat{a}(t) = \Big[-\Dc+\int\hat{\Psi}^\dagger(x,t) U(x)\hat{\Psi}(x,t)\ud x-i\kappa\Big]\hat{a}(t)\\
 + i \eta + i \hat{\xi}(t).
\end{multline}
The parameter $\DC\equiv\omega-\omega_{\mathrm{C}}$ is the cavity detuning,  $\eta$ is the strength of the cavity pumping, $2\kappa$ is the photon escape rate, $U(x)=U_0 \cos^2(k x)$ describes the frequency shift caused by a single atom, with $U_0$ being the strength of this coupling. The noise operator $\hat{\xi}(t)$  describes white noise with the only non-vanishing correlation function
\begin{equation}
\label{eqnoise}
\left<\hat{\xi}(t)\hat{\xi}^\dagger(t')\right>=2\kappa\delta(t-t').
\end{equation}  
The equation of motion of the atomic field operator is
\begin{multline}
\label{eqmoat}
i\hslash\dt\hat{\Psi}(x,t) = \bigg[-\frac{\hslash^2\Delta}{2m}
+\hslash\hat{a}^\dagger(t)\hat{a}(t) U(x)
\bigg]\hat{\Psi}(x,t).
\end{multline}
\end{subequations}
In order to distil the effect that the noise source $\xi$ infiltrates into the matter wave component and induces depletion, the atom-atom s-wave interaction is set artificially to zero.  This is a consistent approach since the cavity-induced depletion is independent of the collisions. The two processes rely on distinct transitions between the Bloch states of the $\lambda/2$ periodic potential  $U(x)$.  The atom-photon interaction does not couple Bloch states with different quasi momenta. We can thus consider wavefunctions being also periodic with $\lambda/2$, the zero quasi-momentum Bloch state representing the entire band. We focus on the depletion due to such `band-to-band' transitions.  By contrast, collisions induce transitions mostly within a Bloch band, and for usual densities this does not influence significantly the band-to-band processes. The validity of the model is restricted to the parameter regime where the Bose gas loaded in the optical lattice created by the cavity field is in the superfluid phase, far from the transition point to the Mott phase \cite{stoferle04,xu06}. In this regime, the small amount of  depletion of the condensate into higher quasi-momentum states due to s-wave scattering can be estimated by the usual formula well-known from  Bogoliubov-theory   \cite{castin}.


The field operators are decomposed according to $\hat{a}(t)=\alpha(t)+\delta\hat{a}(t)$ and $\hat{\Psi}(x,t)=\sqrt{N}\varphi(x,t)+\delta\hat{\Psi}(x,t)$, with $\alpha(t)$ being the coherent component of the radiation field, and similarly $\varphi(x,t)$ stands for the condensate wavefunction normalized to unity, while $N$ is the number of atoms in the condensate. Substituting these into Eqs. \eqref{eqmo} and neglecting those terms which are not pure c-numbers leads to a coupled pair of nonlinear equations  similar in spirit to the Gross-Pitaevskii equations:
\begin{subequations}
\label{gps}
\begin{equation}
\label{gpa}
i\dt\alpha(t) = \big[-\Dc+N \langle U\rangle-i\kappa\big]\alpha(t)
 + i \eta,\\
\end{equation}
\begin{multline}
\label{gpfi}
i\hslash\dt\varphi(x,t) = \bigg[-\frac{\hslash^2\Delta}{2m}+\hslash|\alpha(t)|^2 U(x)
\bigg]\varphi(x,t).
\end{multline} 
\end{subequations}
Via the average $\langle U \rangle=\int \varphi^*(x,t) U(x) \varphi(x,t)\ud x$, the condensate wavefunction $\varphi(x)$ couples into the evolution of the field amplitude. The solution of Eqs. \eqref{gps} yields the possible steady-states of the BEC-cavity system. 

\begin{figure}[t!]
\begin{center}
\includegraphics[scale=.745]{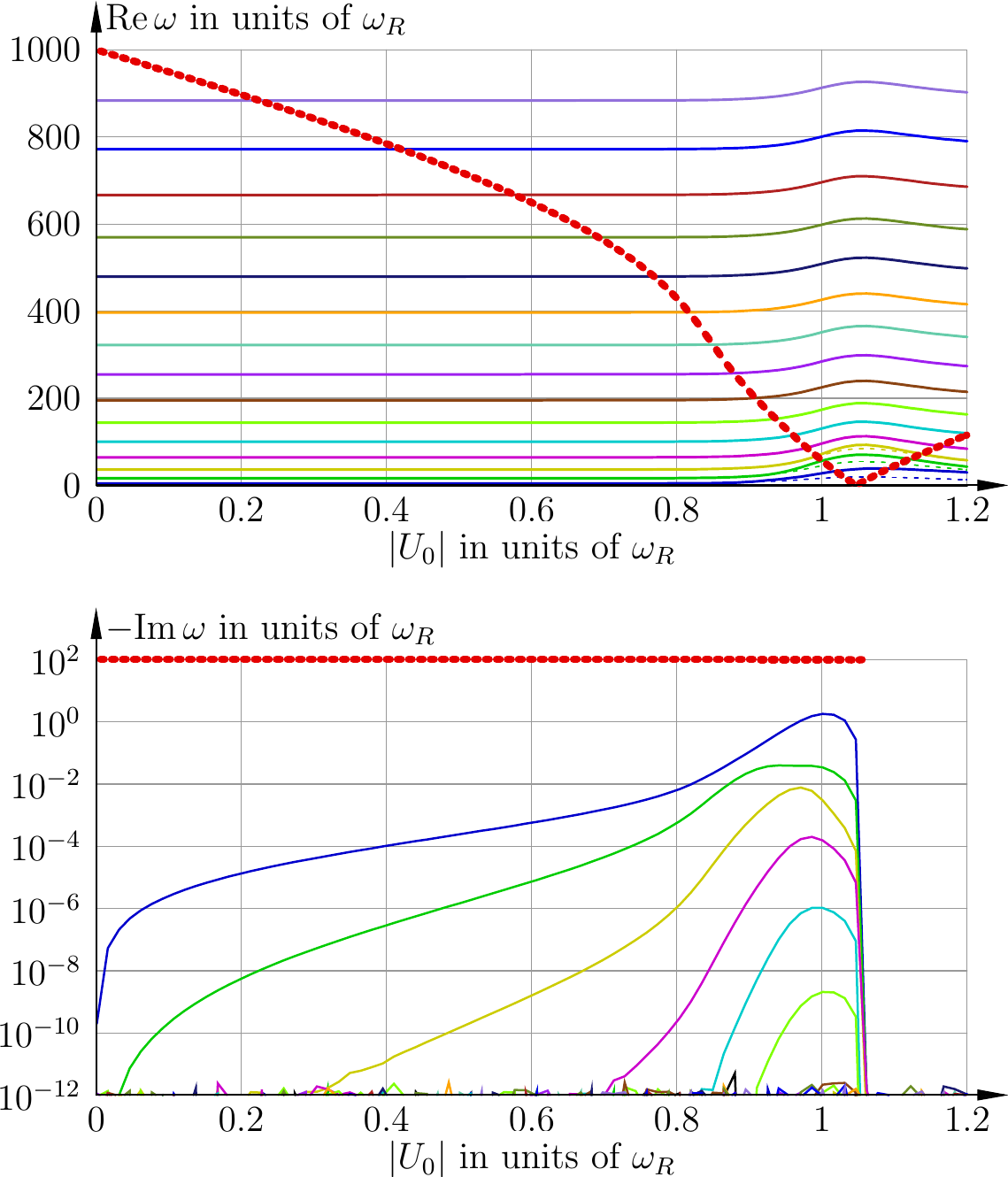}
\caption{(Color online) The real and imaginary parts of the eigenvalues of the linear stability matrix $\mat{M}$ vs. the one atom light shift $U_0$. Only those eigenvalues are plotted which have a nonnegative real part. Every eigenvalue is degenerate at $U_0=0$, and the lift of the degeneracy can be resolved only close to $U_0 \approx 1$ for the low-lying levels, where one of the eigenvalue in the quasi-degenerate pair is plotted by thin, dashed line. These are not coupled to the radiation mode and have vanishing imaginary part.  The highest plotted eigenvalue at $U_0=0$ (thick, dashed line in the top panel) corresponds to a dominantly radiation field excitation; notice that its imaginary part is at $-\kappa$. As $|U_0|$ grows this mode crosses the other excitation energies and reaches zero at $N U_0\approx \DC$. The parameters are $\DC=-1000\,\oR$, $\kappa=100\,\oR$, $\eta=1000\oR$, $N=1000$.}
\label{fig:spectrum}
\end{center}
\end{figure}
The Heisenberg-Langevin equations \eqref{eqmo} can be linearized in the fluctuations $\delta\hat{a}(t)$ and $\delta\hat{\Psi}(x,t)$ around the steady state of Eqs. \eqref{gps}. The procedure is similar to that used in Ref.~\cite{horak01b} and, for a different geometry in Ref.~\cite{nagy08}, however, here we will consider the effect of quantum noise. One obtains a coupled set of differential equations for the fluctuations with couplings to their hermitian adjoints. For a compact notation the fluctuations are arranged in a vector $\R\equiv[\delta\hat{a},\delta\hat{a}^\dagger,\delta\hat{\Psi}(x),\delta\hat{\Psi}^\dagger(x)]$ for which the linearized equation of motion, 
\begin{equation} 
i\dt\R=\mat{M}\R+i\vec{\hat{\xi}}\,,
\end{equation}
where $\vec{\hat{\xi}}=[\hat{\xi},\hat{\xi}^\dagger,0,0]$, and the matrix $\mat{M}$ is 
\begin{equation}
\label{eq:M}
\mat{M}=
\left[
\begin{array}{c c c c}
A&0&\alpha \hat{X}^*&\alpha\hat{X}\\
0&-A^*&-\alpha^* \hat{X}^*&-\alpha^*\hat{X}\\
\alpha^* Y & \alpha Y & \hat{H}_0&0\\
-\alpha^* Y^* & -\alpha Y^* & 0& -\hat{H}_0
\end{array}
\right],
\end{equation}
with $A\equiv-\DC+N\langle U\rangle-i\kappa$, $Y\equiv\sqrt{N}\varphi(x) U(x)$, $\hat{X}f(x)\equiv
\int\varphi(x) U(x) f(x)\ud x$ and $\hat{H}_0\equiv-\hslash\Delta/(2m)+|\alpha|^2U(x)$.  
The key point with respect to the depletion of a BEC is that the matter field fluctuation $\delta\hat{\Psi}(x)$ couples to both $\delta\hat{a}$ and $\delta\hat{a}^\dagger$, this latter being driven by the `noise creation' operator $\xi^\dagger$. Let us also note that the matrix $\mat{M}$ is non-normal, i.e., it doesn't commute with its hermitian adjoint. As a consequence its eigenvectors are not orthogonal to each other, hence the analogy to excess noise in lasers and to the Petermann-factor arises \cite{siegman89,grangier98a,bardroff00a,lamprecht99}.   The stability  can be properly characterized in terms of pseudo-spectra by using the theory of non-normal operators \cite{papoff08a,farrel96}. However, we will restrict this work to the calculation of the depletion of the BEC.

The eigenvalues $\omega_k$ and the corresponding left and right eigenvectors, $\lev{k}$ and $\rev{k}$, respectively, of $\mat{M}$ are calculated numerically. First the coupled Gross-Pitaevskii equations \eqref{gps} are solved on a $200$ point grid with imaginary time propagation. From the wavefunctions, the matrix $\mat{M}$  can be formed and then diagonalized numerically.
The spectrum of fluctuations is exhibited in Fig.~\ref{fig:spectrum}.  At $U_0=0$, the real part (top panel) renders the spectrum of a homogeneous ideal gas. In a large region $0<|U_0|\lesssim 0.8$ the solid lines corresponding to motional excitations are constant, which indicates that the condensate wavefunction remains close to the homogeneous one. There is a critical point  where the condensate pulls the frequency of the cavity mode into resonance with the pump laser frequency ($N \langle U\rangle = \DC$),  marked by that the real part of the eigenvalue of the dominantly cavity-type  excitation (thick, dashed line) crosses zero. At this point, the steady-state wavefunction is localized  in the dipole potential (its depth is about 100 $\omega_R$ for the parameters of the figure). Due to the finite spread of the wavefunction around the antinode, the resonance occurs for $U_0$ slightly shifted from $\Delta_C/N$.

Coupling of the motional modes to the radiation mode is signified by the appearance of non-vanishing imaginary parts (Fig.~\ref{fig:spectrum}, bottom panel). Above the critical point positive imaginary parts appear indicating a dynamical instability of the steady-state solution. This is in accordance with the semiclassical theory of the mechanical forces on atoms in a cavity \cite{domokos03}, which states that the cavity field heats up the motion of atoms in the regime of $\Delta_C-N\langle U\rangle > 0$.  We will disregard this regime and consider only fluctuations around stable steady-state mean field solutions.  Note that the clear maximum of the imaginary parts reflects the semiclassical result that the cavity cooling mechanism is most efficient at half a linewidth below resonance \cite{vukics05b}.


The left- and right eigenvectors of $\mat{M}$  can be used to expand the fluctuation vector $\R$ in terms of ``quasi-normal'' modes:   $\R\equiv\sum_k\hat{\rho}_k\rev{k}$. By use of the orthogonality of the left and right eigenvectors, $(\lev{k},\rev{l})=\delta_{k,l}$, where $(\vec{a},\vec{b})$ is the Euclidean scalar product, the quasi-normal mode amplitude operators are obtained as  $\hat{\rho}_k=(\lev{k},\R)$. They evolve independently
\begin{equation} 
\label{eqmonew}
i\dt\hat{\rho}_k=\omega_k\hat{\rho}_k+i\hat{Q}_k,
\end{equation}
where the projected noise is $\hat{Q}_k\equiv(\lev{k},\vec{\hat{\xi}})$.  With the integration of Eq. \eqref{eqmonew} the time dependence of the fluctuation operators read
\begin{equation} 
\label{tdflu}
\hat{\rho}_k(t)=e^{-i\omega_k t}\hat{\rho}_k(0)+\int_0^te^{-i\omega_k(t-t')}\hat{Q}_k(t')\ud t'.
\end{equation}

Starting from the definition of  the depletion, it can be expressed in terms of the quasi-normal modes as
\begin{multline} 
\delta N(t)=\int\left\langle\delta\hat{\Psi}^\dagger(x,t)\delta\hat{\Psi}(x,t)\right\rangle\ud x\\
=\sum_{k,l}\langle\hat{\rho}_k(t)\hat{\rho}_l(t)\rangle\int r^{(k)}_4(x) r^{(l)}_3(x)\ud x,
\end{multline}
with
\begin{multline}
\label{rhoav}
\left\langle\hat{\rho}_k(t)\hat{\rho}_l(t)\right\rangle=\left\langle\hat{\rho}_k(0)\hat{\rho}_l(0)\right\rangle e^{-i(\omega_k+\omega_l)t}\\
+2\kappa\frac{1-e^{-i(\omega_k+\omega_l)t}}{i(\omega_k+\omega_l)}{l^{(k)}_1}^*{l^{(l)}_2}^*.
\end{multline}
By assuming that we start with a pure BEC at $t=0$, the first term on the right-hand side of Eq. \eqref{rhoav} is zero. The depletion then reads
\begin{multline}
\label{depfin}
\delta N(t)=2\kappa\sum_{k,l}\frac{1-e^{-i(\omega_k+\omega_l)t}}{i(\omega_k+\omega_l)}{l^{(k)}_1}^*{l^{(l)}_2}^*\\
\times\int r^{(k)}_4(x) r^{(l)}_3(x)\ud x.
\end{multline}
Contrary to the case of a conservative system, where quantum depletion is the sum of the individual contributions of the noncondensed one-particle state occupation numbers, here the depletion is the double sum of contributions from pairs of quasi-normal modes ($k$ and $l$). 

The matrix $\mat{M}$ has an important symmetry property, namely $\mat{\Gamma}\mat{M}\mat{\Gamma}=-\mat{M}^*$, originating from the fact that $\R^\dagger=\mat{\Gamma}\R$, with $\mat{\Gamma}$ being the permutation matrix that exchanges the first row with the second and the third row with the fourth one. This property ensures that if $\omega_k$ is an eigenvalue, $-\omega^*_k$ will also be an eigenvalue. Therefore the denominator in Eq. \eqref{depfin} may  become small for a pair of eigenvalues ($\omega_k$ and $\omega_l$) if $\Re\lbrace\omega_k\rbrace=-\Re\lbrace\omega_l\rbrace$, i.e. when the quasi-normal modes $k$ and $l$ are connected with the transformation $\mat{\Gamma}$. Moreover, the smaller the imaginary parts (their dampings), the smaller the denominator.  However, an eigenvalue with a small imaginary part is accompanied by an eigenvector with small ``photon'' components (diminishing $l_1$ and $l_2$ terms). Therefore not all of the eigenvalues with small imaginary parts play a significiant role and the evaluation of the terms remain a numerical task. Nevertheless, the contribution to the depletion from such pairs of quasi-normal modes dominates the double sum.

\begin{figure}[htbp]
\begin{center}
\includegraphics[scale=0.745]{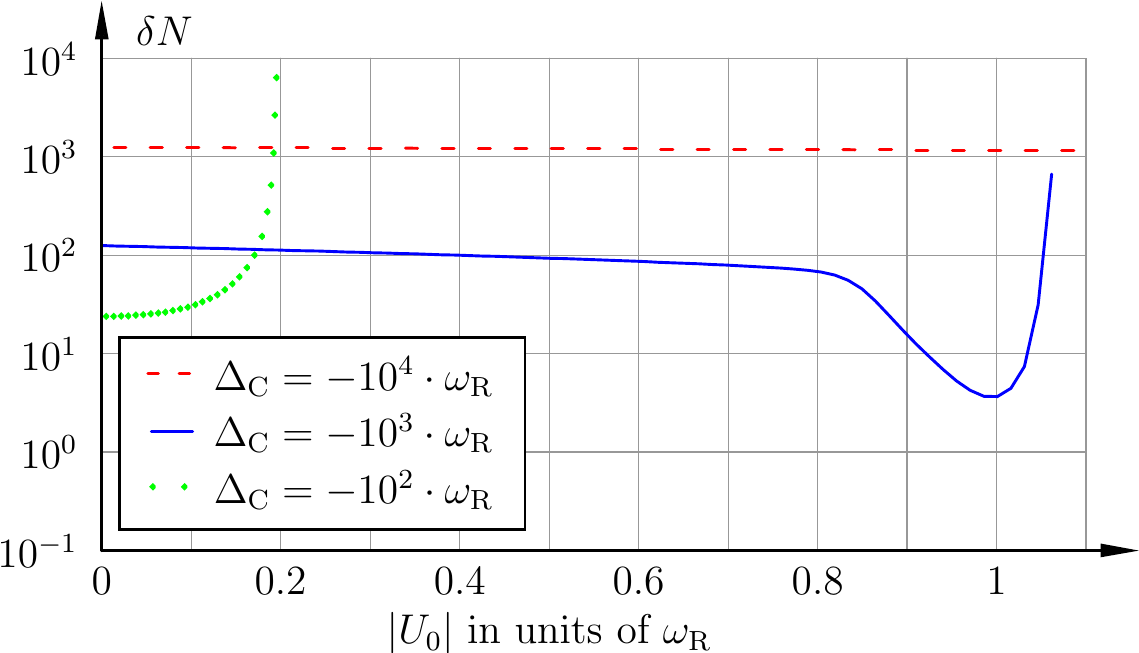}
  \caption{(Color online) The depletion $\delta N$ of the condensate as a function of the coupling constant $U_0$ for various detunings. The parameters are $\kappa=100\,\oR$, $\eta=-\DC$, and $N=1000$. In the plotted range of $U_0$, for smaller detunings (solid and dotted lines) the BEC-cavity system can be resonant with the pump, which leads to divergences. Well below resonance, $N|U_0|\ll \Delta_C$, the depletion depends weakly on $U_0$ (see the plateau of the solid and the dashed lines) and on $N$. In this regime the depletion scales with $\Delta_C/\omega_R$.  }
  \label{cavitypump1}
\end{center}
\end{figure}
The total steady-state depletion in the $t\rightarrow\infty$ limit [the exponential in Eq. \eqref{depfin} is dropped] is shown in Figure \ref{cavitypump1} (semilog scale). The thick solid line corresponds to the parameter set of the spectrum in Fig.\ \ref{fig:spectrum}. The depletion is closely constant for a wide range of $U_0$ up to the resonance regime defined by $|\Delta_C-N\langle U\rangle|\lesssim \kappa$. At resonance the depletion diverges, in accordance with the fact that the dynamical equilibrium of the system becomes unstable. It is interesting that there is a dip slightly below the resonance. We attribute it to the significant change in the shape of the condensate wavefunction, from homogeneous to a strongly localized one, which yields a variation of the overlap factors in the matrix $\mat{M}$.  It is important to note that the finite, large value of depletion in the limit of $U_0\rightarrow0$ was calculated by taking first the $t\rightarrow\infty$ limit of the time dependent result Eq. \eqref{depfin}. However, during a finite measurement time the system does not necessarily  reach the steady state since the relaxation time diverges as $U_0^{-2}$  for small $U_0$.


The actual steady-state value of the depletion  is determined by the ratio of  the cavity mode frequency $\Delta_C$ and the recoil frequency $\omega_R$, i.e., the relation of the energy scales describing optical excitations and atomic motion, respectively. This is illustrated by the other two curves in this figure. Dashed line corresponds to a detuning $\Delta_C$ increased by an order of magnitude, then the depletion is also scaled up by a factor of 10. Obviously, the resonance regime is pushed out of the plotted range and only the initial plateau can be seen. Oppositely, when decreasing $\Delta_C$ by a factor of 10, the initial depletion decreases (dotted line). However, for such a small detuning $\Delta_C\sim \kappa$ the system is in the resonance regime  already for small $U_0$ values, the wide plateau is missing and the divergence of the depletion is exhibited.  As long as the atom-cavity system is far from resonance (on the wide plateau), the depletion $\delta N$ is independent of the atom number $N$. Similarly, the photon number $|\alpha|^2$ is quite irrelevant to the amount of depletion, it influences only the shape of the ground state in the optical potential and thereby some overlap factors in the matrix $\mat{M}$.



In conclusion, we proved that a BEC in a high-Q optical cavity is exposed to excess quantum noise which indicates that the  long-time evolution drives the system into exotic quantum phases unexplored so far.

We acknowledge funding from the National Scientific Fund of Hungary (NF68736, T046129, and T049234).




\end{document}